\documentclass[11pt,epsfig,psfig]{article}
\usepackage{amssymb,amsmath,amsfonts}
\usepackage{graphicx}
\usepackage{graphics}
\usepackage{eepic,epsfig}

\makeatletter

\@addtoreset{equation}{section}

\makeatother

\begin{document}

\title{Whightman function and scalar Casimir densities for a wedge with a
cylindrical boundary}
\author{A. A. Saharian\thanks{%
E-mail: saharyan@www.physdep.r.am } and A. S. Tarloyan \\
Department of Physics, Yerevan State University \\
1 Alex Manoogian St, 375049 Yerevan, Armenia }
\maketitle

\begin{abstract}
Whightman function, the vacuum expectation values of the field
square and the energy-momentum tensor are investigated for a scalar
field inside a wedge with and without a coaxial cylindrical
boundary. Dirichlet boundary conditions are assumed on the bounding
surfaces. The vacuum energy-momentum tensor is evaluated in the
general case of the curvature coupling parameter. Making use of a
variant of the generalized Abel-Plana formula, expectation values
are presented as the sum of two terms. The first one corresponds to
the geometry without a cylindrical boundary and the second one is
induced by the presence of this boundary. The asymptotic behaviour
of the field square, vacuum energy density and stresses near the
boundaries are investigated. The additional vacuum forces acting on
the wedge sides due the presence of the cylindrical boundary are
evaluated and it is shown that these forces are attractive. As a
limiting case, the geometry of two parallel plates perpendicularly
intersected by a third one is analyzed.
\end{abstract}

\bigskip


\bigskip

\section{Introduction}

The influence of boundaries on the vacuum state of a quantum field
leads to interesting physical consequences. The presence of
reflecting boundaries alters the zero-point fluctuations spectrum
and results in the shifts in the vacuum expectation values of
quantities quadratic in the field, such as the energy density and
stresses. In particular, vacuum forces arise acting on
constraining boundaries. The particular features of these forces
depend on the nature of the quantum field, the type of spacetime
manifold, the boundary geometries and the specific boundary
conditions imposed on the field. Since the original work by
Casimir  \cite{Casi48} many theoretical
and experimental works have been done on this problem (see, e.g., \cite%
{Most97,Plun86,Bord01,Milt02} and references therein). However,
there are still difficulties in both interpretation and
renormalization of the Casimir effect. Moreover, the absence of a
complete renormalization procedure, in practice, limits all exact
calculations to the special case of highly-symmetric boundary
configurations (parallel plates, sphere, cylinder) with a specific
background metric. From this point of view the wedge with a
cylindrical boundary is an interesting system, since the geometry is
nontrivial and it includes two dynamical parameters, radius and
angle, for phenomenological purposes. Due to the formal analogy that
exists between a wedge and a straight cosmic string, the
corresponding results can be applied to cosmic strings. In this
paper we will study the vacuum expectation values of the field
square and the energy-momentum tensor for a massless scalar field
with a general curvature coupling parameter inside a wedge of
opening angle $\phi _{0}$ with and without a coaxial cylindrical
boundary assuming the Dirichlet boundary conditions on the
constraining surfaces. The both regions inside and outside the
cylindrical shell are considered. In the discussion below we will
call these regions as region I and region II (see figure
\ref{fig1}). Some most relevant investigations to the present paper
are contained in \cite{jphy,Deutsch,brevikI,brevikII} for a
conformally coupled scalar and electromagnetic fields in a four
dimensional spacetime. The interaction between minimally and
conformally coupled scalar fields and different boundary
configurations constructed from parallel and orthogonal plane
surfaces is studied in \cite{Acto96}. The total Casimir energy of a
semi-circular infinite cylindrical shell with perfectly conducting
walls is considered in \cite{nesterenko} by using the zeta function
technique. For a scalar field with an arbitrary curvature coupling
parameter the vacuum energy density  in the geometry of a wedge with
an
arbitrary opening angle and with a cylindrical boundary is evaluated in \cite%
{Reza02}. This geometry is interesting from the point of view of
general analysis for surface divergences in the expectation values
of local physical observables for boundaries with discontinuities.
The nonsmoothness of the boundary generates additional
contributions to the heat kernel coefficients (see, for instance,
the discussion in \cite{Apps98,Dowk00,Nest03} and references
therein). As in \cite{Reza02}, our method here employs the mode
summation and is based on a variant of the generalized Abel-Plana
formula \cite{Saha87} combined with the point-splitting technique.
This allows to extract from the vacuum expectation values the
parts due to a wedge without a cylindrical shell and to present
the cylindrical parts in terms of strongly convergent integrals.

We have organized the paper as follows. The next section is devoted
to the evaluation of the Wightman function for a massless scalar
field with a general curvature coupling inside a wedge with a
cylindrical boundary. By using the formula for the Wightman
function, in section \ref{sec:Wedgecylabs} we evaluate the vacuum
expectation values of the field square and the energy-momentum
tensor inside a wedge without a cylindrical boundary. In section
\ref{sec:Wedgecyl} we consider the vacuum densities for a wedge with
the cylindrical shell. Formulae for the shell contributions are
derived and the corresponding surface divergences are investigated.
The Wightman function, vacuum expectation values of the field square
and the energy-momentum tensor in the region II are considered in
section \ref{sec:extregion}. As a special case of the considered
geometry, in section \ref{sec:limit} we discuss the limiting case of
two parallel plates perpendicularly intersected by the third plate.
Finally, the results are summarized and discussed in section
\ref{sec:Conc}.

\section{Wightman function inside a wedge}

\label{sec:Wedge}

\begin{figure}[tbph]
\begin{center}
\epsfig{figure=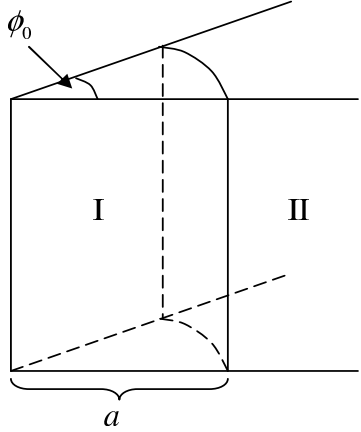, width=5cm, height=6cm}
\end{center}
\caption{Geometry of a wedge with the opening angle $\protect\phi
_0$ and cylindrical boundary of radius $a$.} \label{fig1}
\end{figure}

To describe the geometry of a wedge with the opening angle $\phi
_{0}$ and with the cylindrical boundary of radius $a$ (see figure
\ref{fig1}), we will use the cylindrical coordinates
$(x^{1},x^{2},\ldots ,x^{D})=(r,\phi ,z_{1},\ldots ,z_{N})$,
$N=D-2$, where $D$ is the number of spatial dimensions. Consider a
massless scalar field $\varphi $ with the curvature coupling
parameter $\xi $, satisfying the field equation
\begin{equation}
\left( \nabla ^{i}\nabla _{i}+\xi R\right) \varphi (x)=0,  \label{fieldeq}
\end{equation}%
and obeying Dirichlet boundary condition on the bounding surfaces:
\begin{equation}
\varphi |_{\phi =0}=\varphi |_{\phi =\phi _{0}}=\varphi |_{r=a}=0.
\label{Dirbc}
\end{equation}%
We quote the generalization to other boundary conditions in section \ref%
{sec:Conc}. In (\ref{fieldeq}) $\nabla _{i}$ is the covariant
derivative operator, $R$ is the scalar curvature for the background
spacetime. The values of the curvature coupling parameter $\xi =0$
and $\xi =(D-1)/4D$ correspond to special cases of minimally and
conformally coupled scalars respectively. In this section we
evaluate the positive frequency Wightman function $\langle 0|\varphi
(x)\varphi (x^{\prime })|0\rangle $ in the region I, with $|0\rangle
$ being the amplitude for the corresponding vacuum state. The
response of a particle detector in an arbitrary state of motion is
determined by this function (see, for instance, \cite{Birr82}). In
addition, the vacuum expectation value of the
energy-momentum tensor is expressed in terms of the Wightman function as%
\begin{equation}
\langle 0|T_{ik}(x)|0\rangle =\lim_{x^{\prime }\rightarrow x}\nabla
_{i}\nabla _{k}^{\prime }\langle 0|\varphi (x)\varphi (x^{\prime })|0\rangle
+\left[ \left( \xi -\frac{1}{4}\right) g_{ik}\nabla ^{l}\nabla _{l}-\xi
\nabla _{i}\nabla _{k}\right] \langle 0|\varphi ^{2}(x)|0\rangle .
\label{vevEMTWf}
\end{equation}%
Here we have assumed that the background spacetime is flat and have omitted
the term with the Ricci tensor. By expanding the field operator and using
the standard commutation relations, the Wightman function is presented as
the mode sum%
\begin{equation}
\langle 0|\varphi (x)\varphi (x^{\prime })|0\rangle =\sum_{\mathbf{\alpha }%
}\varphi _{\mathbf{\alpha }}(x)\varphi _{\mathbf{\alpha }}^{\ast
}(x^{\prime }), \label{vevWf}
\end{equation}%
where $\{\varphi _{\mathbf{\alpha }}(x)\}$ is a complete
orthonormal set of positive frequency solutions to the field
equation, satisfying the corresponding boundary conditions,
$\alpha $ is a set of corresponding quantum numbers.

In the region $0\leq r\leq a$ (region I in figure \ref{fig1}), the
eigenfunctions satisfying the boundary conditions (\ref{Dirbc}) on
the wedge sides $\phi =0,\phi _{0}$ have the form
\begin{eqnarray}
\varphi _{\alpha }(x) &=&\beta _{\alpha }J_{qn}(\gamma r)\sin (qn\phi )\exp
\left( i\mathbf{kr}_{\parallel }-i\omega t\right) ,\quad \alpha =(n,\gamma ,%
\mathbf{k}),  \label{eigfunccirc} \\
\omega  &=&\sqrt{\gamma ^{2}+k^{2}},\quad q=\pi /\phi _{0},\quad
-\infty <k_{j}<\infty ,\quad n=1,2,\cdots , \label{qu}
\end{eqnarray}%
where $\mathbf{k}=(k_{1},\ldots ,k_{N})$, $\mathbf{r}_{\parallel
}=(z_{1},\ldots ,z_{N})$, and $J_{l}(z)$ is the Bessel function. The
normalization coefficient $\beta _{\alpha }$ is determined from the standard
Klein-Gordon scalar product with the integration over the region inside the
wedge and is equal to
\begin{equation}
\beta _{\alpha }^{2}=\frac{2}{(2\pi )^{N}\omega \phi _{0}a^{2}J_{qn}^{\prime
2}(\gamma a)}.  \label{betalf}
\end{equation}%
The eigenvalues for the quantum number $\gamma $ are quantized by the
boundary condition (\ref{Dirbc}) on the cylindrical surface $r=a$. From this
condition it follows that the possible values of $\gamma $ are equal to
\begin{equation}
\gamma =\lambda _{n,j}/a,\quad j=1,2,\cdots ,  \label{ganval}
\end{equation}%
where $\lambda _{n,j}$ are the positive zeros of the Bessel function, $%
J_{qn}(\lambda _{n,j})=0$, arranged in ascending order, $\lambda
_{n,j}<\lambda _{n,j+1}$. Substituting the eigenfunctions (\ref{eigfunccirc}%
) into mode sum formula (\ref{vevWf}) with the set of quantum
numbers $\alpha =(n,j,\mathbf{k})$, for the positive frequency
Wightman function one finds
\begin{eqnarray}
\langle 0|\varphi (x)\varphi (x^{\prime })|0\rangle  &=&\int d^{N}\mathbf{k}%
\,e^{i\mathbf{k}(\mathbf{r}_{\parallel }-\mathbf{r}_{\parallel }^{\prime
})}\sum_{n=1}^{\infty }\sin (qn\phi )\sin (qn\phi ^{\prime })  \nonumber \\
&&\times \sum_{j=1}^{\infty }\beta _{\alpha }^{2}J_{qn}(\gamma
r)J_{qn}(\gamma r^{\prime })e^{-i\omega (t-t^{\prime })}|_{\gamma =\lambda
_{n,j}/a}.  \label{Wf1}
\end{eqnarray}%
This formula is not convenient for the further evaluation of the
vacuum expectation values of the field square and the
energy-momentum tensor. This is related to that we do not know the
explicit expressions for the eigenvalues $\lambda _{n,j}$ as
functions on $n$ and $j$, and the summands in the series over $j$
are strongly oscillating functions for large values $j$. In
addition, the expression on the right of (\ref{Wf1}) is divergent
in the coincidence limit and some renormalization procedure is
needed to extract finite result for the vacuum expectation values
of the field square and the energy-momentum tensor.

To obtain an alternative form for the Wightman function we will apply to the
sum over $j$ a variant of the generalized Abel-Plana summation formula \cite%
{Saha87}
\begin{eqnarray}
\sum_{j=1}^{\infty }\frac{2f(\lambda _{n,j})}{\lambda
_{n,j}J_{qn}^{\prime 2}(\lambda _{n,j})\sqrt{\lambda
_{n,j}^{2}+c^{2}}} &=&
\int_{0}^{\infty }\frac{f(z)}{\sqrt{z^{2}+c^{2}}}dz+\frac{\pi }{4}\mathrm{Res%
}_{z=0}\left[ \frac{f(z)Y_{qn}(z)}{\sqrt{z^{2}+c^{2}}J_{qn}(z)}\right] -
\nonumber \\
&&-\frac{1}{\pi
}\int_{0}^{c}dz\,\frac{K_{qn}(z)}{I_{qn}(z)}\frac{e^{-qn\pi
i}f(ze^{\frac{\pi i}{2}})+e^{qn\pi i}f(ze^{-\frac{\pi i}{2}})}{\sqrt{c^{2}-z^{2}}}  \nonumber \\
&&+\frac{i}{\pi }\int_{c}^{\infty }dz\,\frac{K_{qn}(z)}{I_{qn}(z)}\frac{%
e^{-qn\pi i}f(ze^{\frac{\pi i}{2}})-e^{qn\pi i}f(ze^{-\frac{\pi
i}{2}})}{\sqrt{z^{2}-c^{2}}}, \label{sumform1AP}
\end{eqnarray}%
where $Y_{l}(z)$ is the Neumann function, and $I_{l}(z)$, $K_{l}(z)$ are the
Bessel modified functions. This formula is valid for functions $f(z)$
analytic in the right half-plane of the complex variable $z=x+iy$ and
satisfying the conditions
\begin{equation}
|f(z)|<\epsilon (x)e^{c_{1}|y|},\quad c_{1}<2,  \label{sumformcond1}
\end{equation}%
\begin{equation}
f(z)=o(z^{2qn-1}),\quad z\rightarrow 0,  \label{sumformcond2}
\end{equation}%
where $\epsilon (x)\rightarrow 0$ for $x\rightarrow \infty $. By
taking in formula (\ref{sumform1AP}) $qn=1/2$, as a particular
case we receive the Abel-Plana formula (see, for instance,
\cite{Most97}). The generalized Abel-Plana formula was applied
previously to a number of Casimir problems for spherically
\cite{Grig86} and cylindrically \cite{Saha88,saharianI} symmetric
boundaries, and in the braneworld scenarios \cite{Saha05}.

To evaluate the sum over $j$ in (\ref{Wf1}) as a function $f(z)$ we choose
\begin{equation}
f(z)=zJ_{qn}(zr/a)J_{qn}(zr^{\prime }/a)\exp \left[ -i\sqrt{k^{2}+z^{2}/a^{2}%
}(t-t^{\prime })\right] .  \label{ftosum}
\end{equation}%
Using the asymptotic formulae of the Bessel functions for large values of
arguments when $n$ is fixed (see, e.g., \cite{hand}), we can see that for
the function $f(z)$ from (\ref{ftosum}) the conditions (\ref{sumformcond1}),
(\ref{sumformcond2}) are satisfied if $r+r^{\prime }+|t-t^{\prime }|<2a$. In
particular, this is the case in the coincidence limit $t=t^{\prime }$ for
the region under consideration, $r,r^{\prime }<a$. Formula (\ref{sumform1AP}%
) allows to present the Wightman function in the form%
\begin{equation}
\langle 0|\varphi (x)\varphi (x^{\prime })|0\rangle =\langle 0_{w}|\varphi
(x)\varphi (x^{\prime })|0_{w}\rangle +\langle \varphi (x)\varphi (x^{\prime
})\rangle _{a},  \label{Wf2}
\end{equation}%
where%
\begin{eqnarray}
\langle 0_{w}|\varphi (x)\varphi (x^{\prime })|0_{w}\rangle  &=&\frac{1}{%
\phi _{0}}\int \frac{d^{N}\mathbf{k}}{(2\pi )^{N}}e^{i\mathbf{k}(\mathbf{r}%
_{\parallel }-\mathbf{r}_{\parallel }^{\prime })}\int_{0}^{\infty }dz\frac{%
ze^{-i(t-t^{\prime })\sqrt{z^{2}+k^{2}}}}{\sqrt{z^{2}+k^{2}}}  \nonumber \\
&&\times \sum_{n=1}^{\infty }\sin (qn\phi )\sin (qn\phi ^{\prime
})J_{qn}(zr)J_{qn}(zr^{\prime }),  \label{Wf00}
\end{eqnarray}%
and%
\begin{eqnarray}
\langle \varphi (x)\varphi (x^{\prime })\rangle _{a} &=&-\frac{2}{\pi \phi
_{0}}\int \frac{d^{N}\mathbf{k}}{(2\pi )^{N}}e^{i\mathbf{k}(\mathbf{r}%
_{\parallel }-\mathbf{r}_{\parallel }^{\prime })}\int_{k}^{\infty }dz\frac{%
z\cosh \left[ (t-t^{\prime })\sqrt{z^{2}-k^{2}}\right] }{\sqrt{z^{2}-k^{2}}}
\nonumber \\
&&\times \sum_{n=1}^{\infty }\sin (qn\phi )\sin (qn\phi ^{\prime
})I_{qn}(zr)I_{qn}(zr^{\prime })\frac{K_{qn}(za)}{I_{qn}(za)}.  \label{Wfa0}
\end{eqnarray}%
In the limit $a\rightarrow \infty $ for fixed $r,r^{\prime }$, the term $%
\langle \varphi (x)\varphi (x^{\prime })\rangle _{a}$ vanishes and, hence,
the term $\langle 0_{w}|\varphi (x)\varphi (x^{\prime })|0_{w}\rangle $ is
the Wightman function for the wedge without a cylindrical boundary with the
corresponding vacuum state $|0_{w}\rangle $. Consequently, the term $\langle
\varphi (x)\varphi (x^{\prime })\rangle _{a}$ is induced by the presence of
the cylindrical boundary. Hence, the application of the generalized
Abel-Plana formula allowed us to extract from the Wightman function the part
due to the wedge without a cylindrical boundary. For the points away the
cylindrical surface the additional part induced by this surface, formula (%
\ref{Wfa0}), is finite in the coincidence limit and the renormalization is
needed only for the part coming from the term (\ref{Wf00}).

\section{Vacuum expectation values inside a wedge without a cylindrical
boundary}

\label{sec:Wedgecylabs}

In this section we consider the geometry of a wedge without a cylindrical
boundary. Expression (\ref{Wf00}) for the corresponding Wightman function
can be simplified by using the formula \cite{Prud86}%
\begin{equation}
\sideset{}{'}{\sum}_{n=0}^{\infty }\cos nbJ_{qn}(u)J_{qn}(v)=\frac{1}{2q}%
\sum_{n=n_{-}}^{n_{+}}\alpha _{n}J_{0}\left( \sqrt{u^{2}+v^{2}-2uv\cos \frac{%
2\pi n+b}{q}}\right) ,  \label{formn}
\end{equation}%
where $n_{\pm }=\pm \left[ (q\pi \mp b)/(2\pi )\right] $ (square brackets
mean the integer value of the enclosed expression), $\alpha _{n}=1$ for $%
n\neq n_{\pm }$, $\alpha _{n_{\pm }}=1/2$ for $|(q\pi \mp b)/(2\pi
)|=0,1,2,\ldots $, $\alpha _{n_{\pm }}=1$ for $|(q\pi \mp b)/(2\pi )|\neq
0,1,2,\ldots $, and the prime means that the summand with $n=0$ should be
taken with the weight 1/2. This formula is valid for integer values of $%
q=\pi /\phi _{0}\geqslant 1$. The results for other values $q$ are
obtained from the formulae for the renormalized vacuum expectation
values derived below in this section by analytic continuation over
$\phi _{0}$. Making use formula (\ref{formn}), for the sum over
$n$ in formula (\ref{Wf00}) one finds
\begin{equation}
\sum_{n=1}^{\infty }\sin (qn\phi )\sin (qn\phi ^{\prime
})J_{qn}(zr)J_{qn}(zr^{\prime })=\frac{1}{4q}\sum_{j=1}^{2}(-1)^{j+1}%
\sum_{n=n_{j-}}^{n_{j+}}\alpha _{n}J_{0}\left( zu_{j}\right) ,
\label{formn1}
\end{equation}%
where
\begin{eqnarray}
u_{j} &=&\sqrt{r^{2}+r^{\prime 2}-2rr^{\prime }\cos [2n\phi _{0}+\phi
+(-1)^{j}\phi ^{\prime }]},  \label{uj1} \\
n_{j\pm } &=&\pm \left[ \frac{\pi \mp (\phi +(-1)^{j}\phi ^{\prime
})}{2\phi _{0}}\right] . \label{njpm}
\end{eqnarray}%
Now, evaluating the integral over $z$ in (\ref{Wf00}) with the help of
formula \cite{Prud86}%
\begin{equation}
\int_{0}^{\infty }dz\frac{ze^{-i(t-t^{\prime })\sqrt{z^{2}+k^{2}}}}{\sqrt{%
z^{2}+k^{2}}}J_{0}(zu_{j})=\frac{e^{-k\sqrt{u_{j}^{2}-(t-t^{\prime })^{2}}}}{%
\sqrt{u_{j}^{2}-(t-t^{\prime })^{2}}},  \label{intform1}
\end{equation}%
the expression of the Wightman function is presented in the form%
\begin{equation}
\langle 0_{w}|\varphi (x)\varphi (x^{\prime })|0_{w}\rangle =\sum_{j=1}^{2}%
\frac{(-1)^{j+1}}{2(2\pi )^{N+1}}\sum_{n=n_{j-}}^{n_{j+}}\alpha _{n}\int
d^{N}\mathbf{k}e^{i\mathbf{k}(\mathbf{r}_{\parallel }-\mathbf{r}_{\parallel
}^{\prime })}\frac{e^{-k\sqrt{u_{j}^{2}-(t-t^{\prime })^{2}}}}{\sqrt{%
u_{j}^{2}-(t-t^{\prime })^{2}}}.  \label{Wf01}
\end{equation}%
For the integral in this formula one has
\begin{equation}
\int d^{N}\mathbf{k}e^{i\mathbf{k}(\mathbf{r}_{\parallel }-\mathbf{r}%
_{\parallel }^{\prime })}\frac{e^{-k\sqrt{u_{j}^{2}-(t-t^{\prime })^{2}}}}{%
\sqrt{u_{j}^{2}-(t-t^{\prime })^{2}}}=\frac{2^{N}\pi ^{(N-1)/2}\Gamma \left(
\frac{N+1}{2}\right) }{\left[ u_{j}^{2}+|\mathbf{r}_{\parallel }-\mathbf{r}%
_{\parallel }^{\prime }|^{2}-(t-t^{\prime })^{2}\right] ^{(N+1)/2}},
\label{intform1n}
\end{equation}%
with $\Gamma (z)$ being the Euler gamma function. Hence, for the Wightman
function in the geometry of a wedge without a cylindrical boundary one
obtains the formula%
\begin{equation}
\langle 0_{w}|\varphi (x)\varphi (x^{\prime })|0_{w}\rangle =\frac{\Gamma
\left( \frac{D-1}{2}\right) }{4\pi ^{\frac{D+1}{2}}}\sum_{j=1}^{2}%
\sum_{n=n_{j-}}^{n_{j+}}\frac{(-1)^{j+1}\alpha _{n}}{\left[ u_{j}^{2}+|%
\mathbf{r}_{\parallel }-\mathbf{r}_{\parallel }^{\prime }|^{2}-(t-t^{\prime
})^{2}\right] ^{(D-1)/2}}.  \label{Wf02}
\end{equation}%
Note that for the Wightman function of the Minkowski space without
boundaries one has%
\begin{equation}
\langle 0_{M}|\varphi (x)\varphi (x^{\prime })|0_{M}\rangle =\frac{\Gamma
\left( \frac{D-1}{2}\right) }{4\pi ^{\frac{D+1}{2}}\left[ |\mathbf{r}-%
\mathbf{r}^{\prime }|^{2}-(t-t^{\prime })^{2}\right] ^{(D-1)/2}},
\label{WfM0}
\end{equation}%
where $|0_{M}\rangle $ is the amplitude for the vacuum state in the
Minkowski spacetime without boundaries. This function coincides with the $j=1
$, $n=0$ term in formula (\ref{Wf02}). Taking the coincidence limit $%
x^{\prime }\rightarrow x$, for the difference of the vacuum expectation
values of the field square,%
\begin{equation}
\langle \varphi ^{2}(x)\rangle _{\mathrm{ren}}^{(w)}=\langle 0_{w}|\varphi
^{2}(x)|0_{w}\rangle -\langle 0_{M}|\varphi ^{2}(x)|0_{M}\rangle ,
\label{phiren}
\end{equation}%
we find%
\begin{equation}
\langle \varphi ^{2}(x)\rangle _{\mathrm{ren}}^{(w)}=\frac{\Gamma \left(
\frac{D-1}{2}\right) }{(4\pi )^{\frac{D+1}{2}}r^{D-1}}\sum_{j=1}^{2}%
\sideset{}{'}{\sum}_{n=\overline{n}_{j-}}^{\bar{n}_{j+}}\frac{%
(-1)^{j+1}\alpha _{n}}{|\sin \phi _{n}^{(j)}|^{D-1}},  \label{phi2w}
\end{equation}%
where the prime means that the term $j=1$, $n=0$ has to be omitted, and we
use the notations%
\begin{eqnarray}
\phi _{n}^{(j)} &=&n\phi _{0}+\phi \frac{1+(-1)^{j}}{2},  \label{phij} \\
\bar{n}_{j\pm } &=&\pm \left[ \frac{\pi \mp \phi (1+(-1)^{j})}{2\phi _{0}}%
\right] .  \label{nj}
\end{eqnarray}%
In Eq. (\ref{phi2w}), $\alpha _{n}=1$ for $n\neq \overline{n}_{j\pm }$, $%
\alpha _{\overline{n}_{j\pm }}=1/2$ for $|\pi \mp \phi
(1+(-1)^{j})|/2\phi _{0}=0,1,2,\ldots $, $\alpha
_{\overline{n}_{j\pm }}=1$ for $|\pi \mp \phi (1+(-1)^{j})|/2\phi
_{0}\neq 0,1,2,\ldots $. In the case $D=3$, by using the formula
\begin{equation}
\sum_{n=0}^{m-1}\sec ^{2}\left( x+\frac{n\pi }{m}\right) =m^{2}\csc
^{2}\left( mx+\frac{m\pi }{2}\right) ,  \label{sum3}
\end{equation}%
for the renormalized vacuum expectation value of the field square
one finds \cite{Deutsch}%
\begin{equation}
\langle \varphi ^{2}(x)\rangle _{\mathrm{ren}}^{(w)}=\frac{%
q^{2}-1-3q^{2}\csc ^{2}\left( q\phi \right) }{48\pi ^{2}r^{2}}.
\label{phi2D3}
\end{equation}

Near the wedge boundaries $\phi =\phi _{m}$, $m=0,1$ ($\phi _{1}=0$) the
main contribution in (\ref{phi2w}) comes from the terms $j=2$, $n=0$ and $%
n=-1$ for $m=0$ and $m=1$ respectively, and the renormalized vacuum
expectation value of the field square diverges with the leading behaviour%
\begin{equation}
\langle \varphi ^{2}(x)\rangle _{\mathrm{ren}}^{(w)}=-\frac{\Gamma \left(
\frac{D-1}{2}\right) }{(4\pi )^{\frac{D+1}{2}}\left( r|\phi -\phi
_{m}|\right) ^{D-1}},\quad \phi \rightarrow \phi _{m}.  \label{phi2wedge}
\end{equation}%
The surface divergences in renormalized vacuum expectation values of the
local physical observables are well known in quantum field theory with
boundaries and result from the idealization of the boundaries as perfectly
smooth surfaces which are perfect reflectors at all frequencies. These
divergences are investigated in detail for various types of fields and
general shape of smooth boundary \cite{Deutsch,kennedy}. Near the smooth
boundary the leading divergence in the field square varies as $(D-1)$th
power of the distance from the boundary. It seems plausible that such
effects as surface roughness, or the microstructure of the boundary on small
scales (the atomic nature of matter for the case of the electromagnetic
field \cite{Cand80}) can introduce a physical cutoff needed to produce
finite values of surface quantities. An alternative mechanism for
introducing a cutoff which removes singular behavior on boundaries is to
allow the position of the boundary to undergo quantum fluctuations \cite%
{Ford98}. Such fluctuations smear out the contribution of the high frequency
modes without the need to introduce an explicit high frequency cutoff.

Now we turn to the vacuum expectation values of the energy-momentum tensor.
By making use formula (\ref{vevEMTWf}), for the non-zero components one
obtains (no summation over $i$)%
\begin{eqnarray}
\langle T_{i}^{i}\rangle _{\mathrm{ren}}^{(w)} &=&-\frac{\Gamma \left( \frac{%
D+1}{2}\right) }{2^{D+2}\pi ^{\frac{D+1}{2}}r^{D+1}}\sum_{j=1}^{2}%
\sideset{}{'}{\sum}_{n=\bar{n}_{j-}}^{\bar{n}_{j+}}\frac{(-1)^{j+1}\alpha
_{n}f_{jn}^{(i)}}{|\sin \phi _{n}^{(j)}|^{D+1}},  \label{Tiiw} \\
\langle T_{2}^{1}\rangle _{\mathrm{ren}}^{(w)} &=&-\frac{D(\xi -\xi
_{c})\Gamma \left( \frac{D+1}{2}\right) }{2^{D}\pi ^{\frac{D+1}{2}}r^{D}}%
\sum _{n=\bar{n}_{2-}}^{\bar{n}_{2+}}\frac{\alpha _{n}\sin \phi
_{n}^{(2)}\cos \phi _{n}^{(2)}}{|\sin \phi
_{n}^{(2)}|^{D+1}},\label{Tiiw21}
\end{eqnarray}%
where $i=0,1,\ldots ,D$, and we use the following notations%
\begin{eqnarray}
f_{jn}^{(i)} &=&1+\left( 4\xi -1\right) \left[ (D-1)\delta _{j1}\sin
^{2}\phi _{n}^{(j)}+D\delta _{j2}\right] ,\quad i=0,3,\ldots ,D,  \label{fji}
\\
f_{jn}^{(1)} &=&f_{jn}^{(0)}-4D(\xi -\xi _{c})\sin ^{2}\phi _{n}^{(j)},
\label{fji1} \\
f_{jn}^{(2)} &=&D\left[ 4\sin ^{2}\phi _{n}^{(j)}\left( \xi -\xi _{c}\delta
_{j2}\right) -\delta _{j1}\right] .  \label{fji2}
\end{eqnarray}%
In the case $\phi _{0}=\pi /2$ and for minimally and conformally
coupled scalar fields, it can be checked that from formulae
(\ref{Tiiw}) and (\ref{Tiiw21}), after the transformation from
cylindrical coordinates to the cartesian ones, as a special case we
obtain the result derived in \cite{Acto96}. For a scalar field with
a general curvature coupling parameter the vacuum energy-momentum
tensor is non-diagonal. For the conformally coupled scalar field
this tensor is diagonal and does not depend on the angular
coordinate $\phi $. From the continuity equation $\nabla
_{k}T_{i}^{k}=0$ for the energy-momentum tensor one has the
following
relations for the components:%
\begin{eqnarray}
\partial _{r}\left( rT_{2}^{1}\right) +r\partial _{\phi }T_{2}^{2} &=&0,
\label{conteq1} \\
\partial _{r}\left( rT_{1}^{1}\right) +r\partial _{\phi }T_{1}^{2}
&=&T_{2}^{2}.  \label{conteq2}
\end{eqnarray}%
In the case of a conformally coupled scalar field we have an additional zero
trace condition. It can be checked that vacuum expectation values (\ref{Tiiw}%
), (\ref{Tiiw21}) satisfy equations (\ref{conteq1}), (\ref{conteq2}).

For a non-conformally coupled field vacuum expectation values (\ref{Tiiw})
diverge on the boundaries $\phi =\phi _{m}$. For the points near these
boundaries the leading terms in the corresponding asymptotic expansions are
given by relations (no summation over $i$, $i=0,3,\ldots ,D$)%
\begin{equation}
\langle T_{i}^{i}\rangle _{\mathrm{ren}}^{(w)}\approx \frac{\langle
T_{2}^{2}\rangle _{\mathrm{ren}}^{(w)}}{(\phi -\phi _{m})^{2}}\approx \frac{%
-\langle T_{2}^{1}\rangle _{\mathrm{ren}}^{(w)}}{r(\phi -\phi _{m})}\approx
\frac{D(\xi -\xi _{c})\Gamma \left( \frac{D+1}{2}\right) }{2^{D}\pi ^{\frac{%
D+1}{2}}\left( r|\phi -\phi _{m}|\right) ^{D+1}}.  \label{Tiinearwedge}
\end{equation}%
For the points away from the edge $r=0$, this leading behavior is the same
as that for the geometry of a single plate.

In the most important case $D=3$, by using formula (\ref{sum3}), for the
components of the renormalized energy-momentum tensor we find%
\begin{eqnarray}
\langle T_{0}^{0}\rangle _{\mathrm{ren}}^{(w)} &=&\langle T_{3}^{3}\rangle _{%
\mathrm{ren}}^{(w)}=\frac{1}{32\pi ^{2}r^{4}}\left\{ \frac{1-q^{4}}{45}+%
\frac{8}{3}\left( 1-q^{2}\right) (\xi -\xi _{c})\right.   \notag \\
&&\left. +12\frac{(\xi -\xi _{c})q^{2}}{\sin ^{2}(q\phi )}\left[ \frac{q^{2}%
}{\sin ^{2}(q\phi )}-\frac{2}{3}q^{2}+\frac{2}{3}\right] \right\} ,
\label{TikD3} \\
\langle T_{1}^{1}\rangle _{\mathrm{ren}}^{(w)} &=&\frac{1}{32\pi ^{2}r^{4}}%
\left\{ \frac{1-q^{4}}{45}-\frac{4}{3}(1-q^{2})(\xi -\xi _{c})\right.
\notag \\
&&\left. +12\frac{(\xi -\xi _{c})q^{2}}{\sin ^{2}\left( q\phi \right) }\left[
\frac{q^{2}}{\sin ^{2}\left( q\phi \right) }-\frac{2}{3}q^{2}-\frac{1}{3}%
\right] \right\} ,  \label{TikD311} \\
\langle T_{2}^{1}\rangle _{\mathrm{ren}}^{(w)} &=&-\frac{3(\xi -\xi _{c})}{%
8\pi ^{2}r^{3}}\frac{q^{3}\cos \left( q\phi \right) }{\sin ^{3}\left( q\phi
\right) },  \label{TikD321} \\
\langle T_{2}^{2}\rangle _{\mathrm{ren}}^{(w)} &=&\frac{1}{8\pi ^{2}r^{4}}%
\left[ \frac{q^{4}-1}{60}+(\xi -\xi _{c})\left( 1-q^{2}+\frac{3q^{2}}{\sin
^{2}\left( q\phi \right) }\right) \right] .  \label{TikD322}
\end{eqnarray}%
For a conformally coupled scalar field this tensor coincides with the result
previously obtained in literature \cite{jphy,Deutsch}. The corresponding
vacuum forces acting on the wedge sides are determined by the effective
pressure $-\langle T_{2}^{2}\rangle _{\mathrm{ren}}^{(w)}$. These forces are
attractive for the wedge with $q>1$ and are repulsive for $q<1$. Formula (%
\ref{TikD3}) for the component $\langle T_{0}^{0}\rangle _{\mathrm{ren}%
}^{(w)}$ is derived in \cite{Reza02} by a different way. For the
electromagnetic field in $D=3$ the vacuum expectation values of the
energy-momentum tensor are considered in \cite{Deutsch,brevikI} for
a perfectly conducting wedge and in \cite{brevikII} for a dielectric
wedge. In both cases the vacuum energy-momentum tensor is uniform.
This is a consequence of the conformal invariance of the
electromagnetic field.

\section{Field square and the vacuum energy-momentum tensor in the region~I}

\label{sec:Wedgecyl}

We now turn to the geometry of a wedge with additional cylindrical boundary
of radius $a$. Taking the coincidence limit $x^{\prime }\rightarrow x$ in
formula (\ref{Wf2}) for the Wightman function and integrating over $\mathbf{k%
}$ with the help of the formula \cite{saharianI}
\begin{equation}
\int d^{N}\mathbf{k}\int_{k}^{\infty }\frac{k^{s}g(z)dz}{\sqrt{z^{2}-k^{2}}}=%
\frac{\pi ^{N/2}}{\Gamma (N/2)}B\left( \frac{N+s}{2},\frac{1}{2}\right)
\int_{0}^{\infty }dz\,z^{N+s-1}g(z),  \label{intk}
\end{equation}%
where $B(x,y)$ is the Euler beta function, the vacuum expectation value of
the field square is presented as a sum of two terms%
\begin{equation}
\langle 0|\varphi ^{2}|0\rangle =\langle 0_{w}|\varphi ^{2}|0_{w}\rangle
+\langle \varphi ^{2}\rangle _{a},  \label{phi2a}
\end{equation}%
where the part induced by the cylindrical boundary is given by formula%
\begin{equation}
\langle \varphi ^{2}\rangle _{a}=-\frac{2^{3-D}\pi ^{\frac{1-D}{2}}}{\Gamma
\left( \frac{D-1}{2}\right) a^{D-1}\phi _{0}}\sum_{n=1}^{\infty }\sin
^{2}(qn\phi )\int_{0}^{\infty }dz\,z^{D-2}\frac{K_{qn}(z)}{I_{qn}(z)}%
I_{qn}^{2}(zr/a).  \label{phi2a1}
\end{equation}%
Note that this part vanishes at the wedge sides $\phi =\phi _{m}$, $0\leq r<a
$. Near the edge $r=0$ the main contribution into $\langle \varphi
^{2}\rangle _{a}$ comes from the summand $n=1$ and one has%
\begin{equation}
\langle \varphi ^{2}\rangle _{a}\approx \frac{4\Gamma ^{-2}(q+1)\sin
^{2}(q\phi )\mathcal{F}_{D}(q)}{(4\pi )^{\frac{D-1}{2}}\Gamma \left( \frac{%
D-1}{2}\right) a^{D-1}\phi _{0}}\left( \frac{r}{2a}\right) ^{2q}
\label{phi2ar0}
\end{equation}%
for $r\ll a$, with the notation%
\begin{equation}
\mathcal{F}_{D}(q)=\int_{0}^{\infty }dz\,z^{D+2q-2}\frac{K_{q}(z)}{I_{q}(z)}.
\label{IDq}
\end{equation}

The part $\langle \varphi ^{2}\rangle _{a}$ diverges on the cylindrical
surface $r=a$. Near this surface the main contribution into (\ref{phi2a1})
comes from large values $n$. Introducing a new integration variable $%
z\rightarrow nqz$, replacing the Bessel modified functions by their uniform
asymptotic expansions for large values of the order (see, for instance, \cite%
{hand}), and expanding over $a-r$, for $|\phi -\phi _{m}|\gg 1-r/a$ to the
leading order one finds%
\begin{eqnarray}
\langle \varphi ^{2}\rangle _{a} &\approx &-\frac{(q/2a)^{D-1}}{\pi ^{\frac{%
D+1}{2}}\Gamma \left( \frac{D-1}{2}\right) }\int_{0}^{\infty }dz\frac{z^{D-1}%
}{\sqrt{1+z^{2}}}\sum_{n=1}^{\infty }n^{D-2}e^{-2nq(1-r/a)\sqrt{1+z^{2}}}
 \nonumber \\
&\approx &-\frac{\Gamma \left( \frac{D-1}{2}\right) }{(4\pi )^{\frac{D+1}{2}%
}(a-r)^{D-1}}.  \label{Phi2neara}
\end{eqnarray}%
This leading behavior is the same as that for a cylindrical surface of
radius $a$.

Similarly, the vacuum expectation value of the energy-momentum tensor for
the situation when the cylindrical boundary is present is written in the form%
\begin{equation}
\langle 0|T_{ik}|0\rangle =\langle 0_{w}|T_{ik}|0_{w}\rangle +\langle
T_{ik}\rangle _{a},  \label{Tika}
\end{equation}%
where $\langle T_{ik}\rangle _{a}$ is induced by the cylindrical boundary.
This term is obtained from the corresponding part in the Wightman function, $%
\langle \varphi (x)\varphi (x^{\prime })\rangle _{a}$, acting by the
appropriate differential operator and taking the coincidence limit [see
formula (\ref{vevEMTWf})]. For the points away from the cylindrical surface
this limit gives a finite result. For the corresponding components of the
energy-momentum tensor one obtains (no summation over $i$)%
\begin{eqnarray}
\langle T_{i}^{i}\rangle _{a} &=&\frac{(4\pi )^{-\frac{D-1}{2}}}{\Gamma
\left( \frac{D-1}{2}\right) a^{D+1}\phi _{0}}\sum_{n=1}^{\infty
}\int_{0}^{\infty }dzz^{D}\frac{K_{qn}(z)}{I_{qn}(z)}  \nonumber \\
&&\times \left\{
a_{i,qn}^{(+)}[I_{qn}(y)]-a_{i,qn}^{(-)}[I_{qn}(y)]\cos (2qn\phi
)\right\}
,\quad y=\frac{zr}{a},  \label{Tiia} \\
\langle T_{2}^{1}\rangle _{a} &=&\frac{2(4\pi )^{-\frac{D-1}{2}}}{\Gamma
\left( \frac{D-1}{2}\right) a^{D}\phi _{0}}\sum_{n=1}^{\infty }qn\sin
(2qn\phi )\int_{0}^{\infty }dz\,z^{D-1}\frac{K_{qn}(z)}{I_{qn}(z)}  \nonumber \\
&&\times I_{qn}(y)\left[ \frac{2\xi }{y}I_{qn}(y)+(1-4\xi )I_{qn}^{\prime
}(y)\right] ,  \label{Tiia21}
\end{eqnarray}%
with the notations%
\begin{eqnarray}
a_{i,l}^{(\pm )}[g(y)] &=&(4\xi -1)\left\{ g^{\prime 2}(y)+g^{2}(y)
\left[ 1\pm \frac{l^{2}}{y^{2}}+\frac{2}{(D-1)(4\xi -1)}\right]
\right\} ,
\label{ajpm} \\
a_{1,l}^{(\pm )}[g(y)] &=&g^{\prime 2}(y)+\frac{4\xi }{y}%
g(y)g^{\prime }(y)-g^{2}(y)\left\{ 1\pm \left[ 1-4\xi (1\mp 1)%
\right] \frac{l^{2}}{y^{2}}\right\} ,  \label{ajpm1} \\
a_{2,l}^{(\pm )}[g(y)] &=&\left( 4\xi -1\right) \left[ g^{\prime
2}(y)+g^{2}(y)\right] -\frac{4\xi }{y}g(y)g^{\prime }(y)+\frac{%
l^{2}}{y^{2}}g^{2}(y)\left( 4\xi \pm 1\right) ,  \label{ajpm2}
\end{eqnarray}%
for a given function $g(y)$, $i=0,3,\ldots ,D$. Formula (\ref{Tiia})
for the energy density ($i=0$) was previously derived in
\cite{Reza02}. In accordance with the problem symmetry, the
expressions for the diagonal components are invariant under the replacement $%
\phi \rightarrow \phi _{0}-\phi $, and the off-diagonal component $\langle
T_{2}^{1}\rangle _{a}$ changes the sign under this replacement. Note that
the latter vanishes on the wedge sides $\phi =\phi _{m}$, $0\leq r<a$ and
for $\phi =\phi _{0}/2$. On the wedge sides one has $\cos (2qn\phi )=1$ and
for the diagonal components of the energy-momentum tensor we obtain (no
summation over $i$)%
\begin{equation}
\langle T_{i}^{i}\rangle _{a}|_{\phi =\phi _{m}}=\frac{2^{2-D}\pi ^{\frac{5-D%
}{2}}A_{i}}{\Gamma \left( \frac{D-1}{2}\right) a^{D-1}r^{2}\phi _{0}^{3}}%
\sum_{n=1}^{\infty }n^{2}\int_{0}^{\infty }dz\,z^{D-2}\frac{K_{qn}(z)}{%
I_{qn}(z)}I_{qn}^{2}(zr/a),  \label{Tiionphim}
\end{equation}%
where $A_{i}=4\xi -1$, $i=0,1,3,\ldots ,D$, $A_{2}=1$. In particular, the
additional vacuum effective pressure in the direction perpendicular to the
wedge sides, $p_{a}=-\langle T_{2}^{2}\rangle _{a}|_{\phi =\phi _{m}}$, does
not depend on the curvature coupling parameter and is negative for all
values $0<r<a$. This means that the vacuum forces acting on the wedge sides $%
\phi =\phi _{m}$ due to the presence of the cylindrical boundary are
attractive. The corresponding vacuum stresses in the directions parallel to
the wedge sides are isotropic and the energy density is negative for both
minimally and conformally coupled scalars. It can be checked that
expectation values (\ref{Tiia}), (\ref{Tiia21}) satisfy equations (\ref%
{conteq1}), (\ref{conteq2}) and, hence, the continuity equation for the
energy-momentum tensor. Aiming to compare with the result for the
energy-momentum tensor of a cylindrical shell with the radius $a$ let us
write down the corresponding formula, which is obtained from the general
result of \cite{saharianI} and has the form
\begin{equation}
\langle T_{i}^{k}\rangle _{\mathrm{ren}}^{\mathrm{cyl}}=\frac{2(4\pi )^{-%
\frac{D+1}{2}}\delta _{i}^{k}}{a^{D+1}\Gamma \left( \frac{D-1}{2}\right) }%
\sum_{n=-\infty }^{+\infty }\int_{0}^{\infty }dz\,z^{D}\frac{K_{n}(z)}{%
I_{n}(z)}a_{i,n}^{(+)}[I_{n}(zr/a)] ,  \label{cyldens}
\end{equation}%
with the same notations as in (\ref{ajpm})-(\ref{ajpm2}).

For $0<r<a$ the cylindrical parts (\ref{Tiia}) and (\ref{Tiia21}) are finite
for all values $0\leq \phi \leq \phi _{0}$, including the wedge sides. The
divergences on these sides are included in the first term on the right-hand
side of (\ref{Tika}) corresponding to the case without cylindrical boundary.
Near the edge $r=0$ the main contribution into the boundary part (\ref{Tiia}%
) comes from the summand with $n=1$ and one has
\begin{eqnarray}
\langle T_{i}^{i}\rangle _{a} &\approx &\frac{q\left[
B_{i}^{(+)}-B_{i}^{(-)}\cos (2q\phi )\right] \mathcal{F}_{D}(q)}{(4\pi )^{%
\frac{D+1}{2}}a^{D+1}\Gamma \left( \frac{D-1}{2}\right) \Gamma ^{2}(q)}%
\left( \frac{r}{2a}\right) ^{2q-2},  \label{Tiir0} \\
\langle T_{2}^{1}\rangle _{a} &\approx &\frac{\left[ 2\xi +(1-4\xi
)q\right] \sin (2q\phi )\mathcal{F}_{D}(q)}{\pi (4\pi
)^{\frac{D-1}{2}}a^{D}\Gamma \left( \frac{D-1}{2}\right) \Gamma
^{2}(q)}\left( \frac{r}{2a}\right) ^{2q-1},  \label{Tii21r0}
\end{eqnarray}%
with the notations%
\begin{eqnarray}
B_{i}^{(\pm )} &=&(4\xi -1)(1\pm 1),\quad i=0,3,\ldots ,D,  \label{Bipm} \\
B_{1}^{(\pm )} &=&4\xi (1/q-1\pm 1)\mp 1+1,\quad B_{2}^{(\pm
)}=4\xi (2-1/q)\pm 1-1,  \nonumber
\end{eqnarray}%
and the function $\mathcal{F}_{D}(q)$ is defined by formula (\ref{IDq}).

The boundary part $\left\langle T_{i}^{k}\right\rangle _{a}$ diverges on the
cylindrical surface $r=a$. Introducing a new integration variable $%
z\rightarrow nqz$ and by taking into account that near the surface $r=a$ the
main contribution comes from the large values of $n$ we can replace the
Bessel modified functions by their uniform asymptotic expansions for large
values of the order. Expanding over $a-r$, to the leading order for the
diagonal components one finds%
\begin{eqnarray}
\langle T_{i}^{i}\rangle _{a} &\approx &\frac{(4\pi )^{-\frac{D-1}{2}}}{%
2\Gamma \left( \frac{D-1}{2}\right) a^{D+1}\phi _{0}}\sum_{n=1}^{\infty
}(qn)^{D}\int_{0}^{\infty }dz\,z^{D-2}(1+z^{2})^{1/2} \nonumber  \\
&&\times e^{-2nq(1-r/a)\sqrt{1+z^{2}}}\left[ A_{i}^{(+)}(z)-A_{i}^{(-)}(z)%
\cos (2qn\phi )\right] .\label{intsum3}
\end{eqnarray}%
where
\begin{eqnarray}
A_{0}^{(\pm )}(y) &=&(4\xi -1)\left\{ 1+\frac{1}{1+y^{2}}\left[ y^{2}\pm 1+%
\frac{2y^{2}}{(D-1)(4\xi -1)}\right] \right\} ,  \label{bepm} \\
A_{1}^{(\pm )}(y) &=&1-\frac{1}{1+y^{2}}\left[ y^{2}\pm 1+4\xi (1\mp 1)%
\right] ,\label{bepm1} \\
A_{2}^{(\pm )}(y) &=&4\xi -1+\frac{1}{1+y^{2}}\left[ y^{2}(4\xi
-1)+4\xi \pm 1\right] ,\label{bepm2}
\end{eqnarray}%
On the wedge sides one has $\cos (2qn\phi )=1$ and this yields
\begin{equation}
\langle T_{i}^{i}\rangle _{a}|_{\phi =\phi _{m}}\approx \frac{(4\pi )^{-%
\frac{D-1}{2}}A_{i}}{\Gamma \left( \frac{D-1}{2}\right) a^{D+1}\phi _{0}}%
\sum_{n=1}^{\infty }(qn)^{D}\int_{0}^{\infty }\frac{dz\,z^{D-2}}{\sqrt{%
1+z^{2}}}e^{-2nq(1-r/a)\sqrt{1+z^{2}}},  \label{intsum4}
\end{equation}%
where the coefficients $A_{i}$ are defined in the paragraph after formula (%
\ref{Tiionphim}). Summing over $n$, to the leading order over $(a-r)^{-1}$%
one finds
\begin{equation}
\langle T_{i}^{i}\rangle _{a}|_{\phi =\phi _{m}}\approx \frac{A_{i}\Gamma
\left( \frac{D+1}{2}\right) }{2(4\pi )^{\frac{D+1}{2}}(a-r)^{D+1}},\quad
r\rightarrow a.  \label{T00asra1}
\end{equation}%
It can be seen that for the non-diagonal component to the leading order one
has $\langle T_{2}^{1}\rangle _{a}\sim (a-r)^{-D}$.

\begin{figure}[tbph]
\begin{center}
\begin{tabular}{cc}
\epsfig{figure=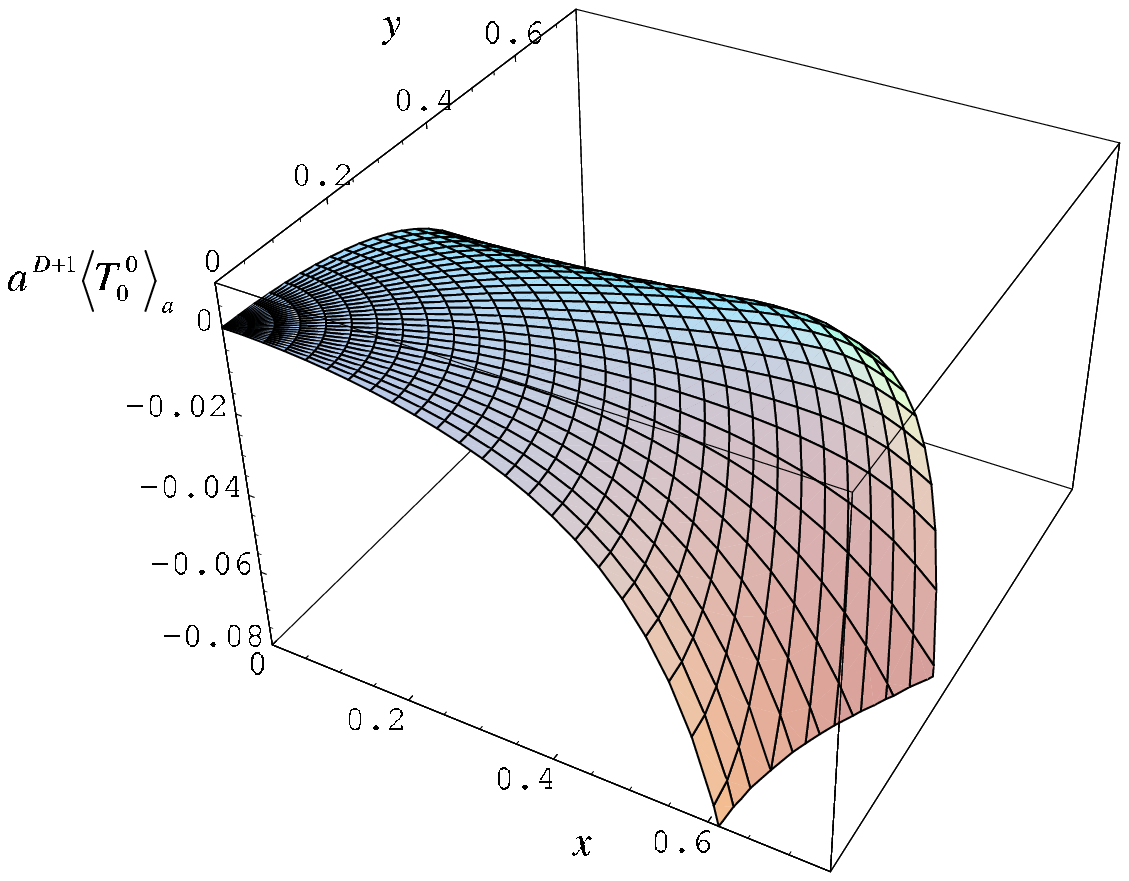, width=7cm, height=6cm} & \quad %
\epsfig{figure=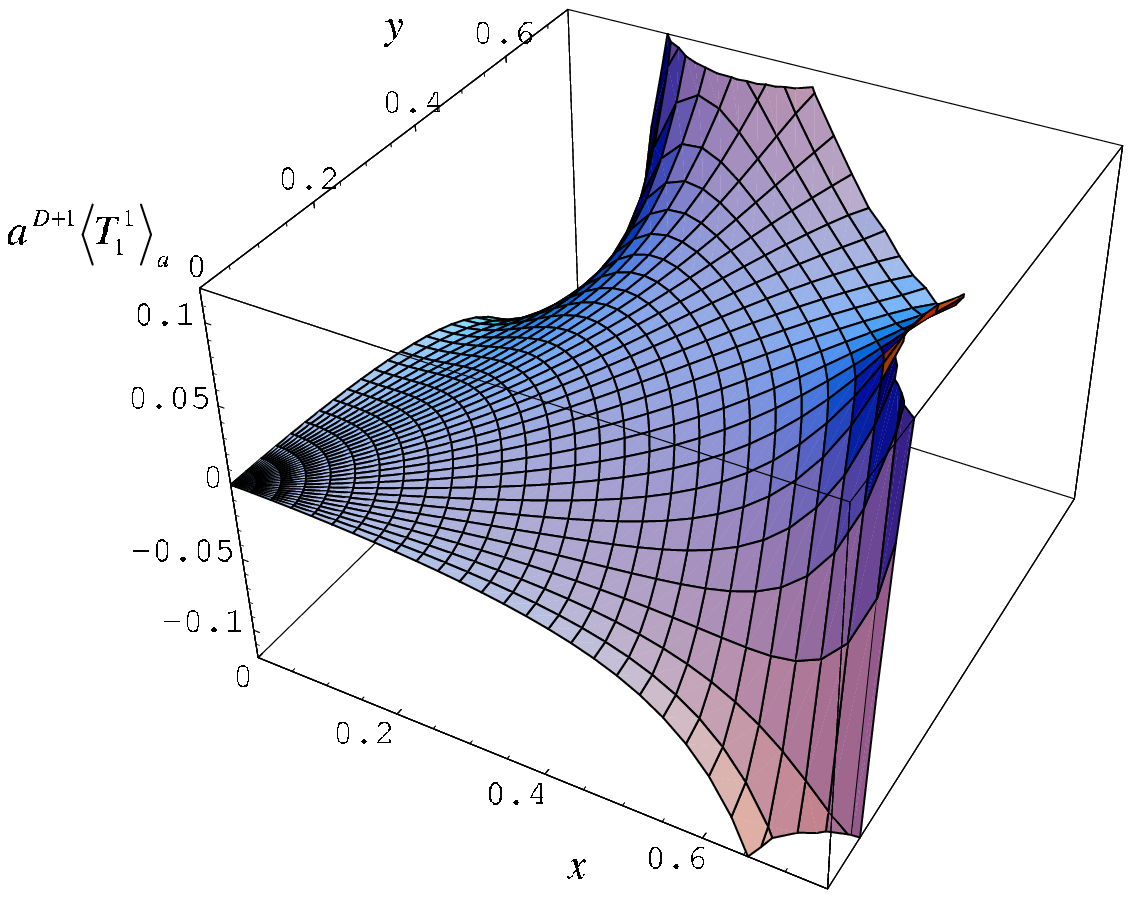, width=7cm, height=6cm}%
\end{tabular}%
\end{center}
\caption{Dependence of the energy density and radial stress for a
$D=3$ conformally coupled scalar field, induced by a cylindrical
boundary of radius $a$, on coordinates $r$ and $\protect\phi $ for
the wedge with the opening angle $\protect\phi _0=\protect\pi /2$.
The variables on the axes are $x=(r/a)\cos \protect\phi $ and
$y=(r/a)\sin \phi $.} \label{fig2}
\end{figure}

\begin{figure}[tbph]
\begin{center}
\begin{tabular}{cc}
\epsfig{figure=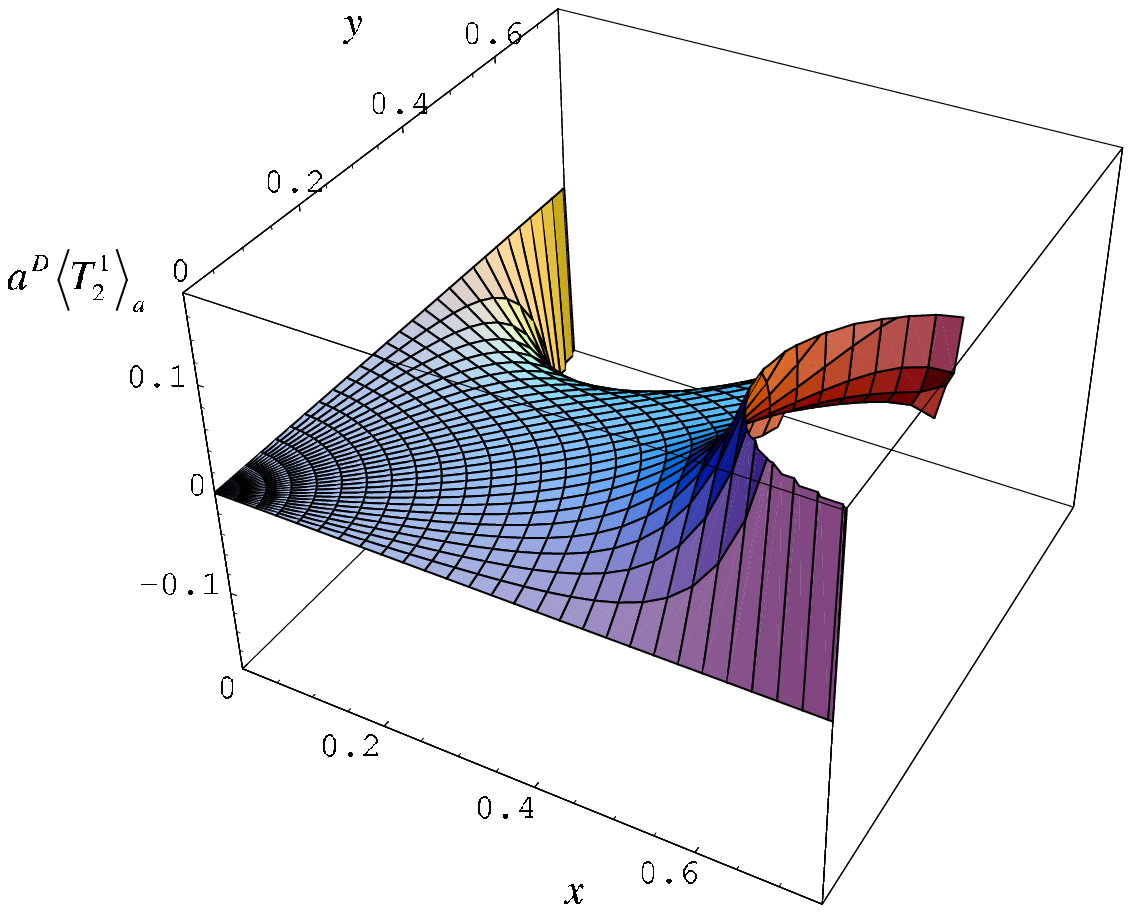, width=7cm, height=6cm} & \quad %
\epsfig{figure=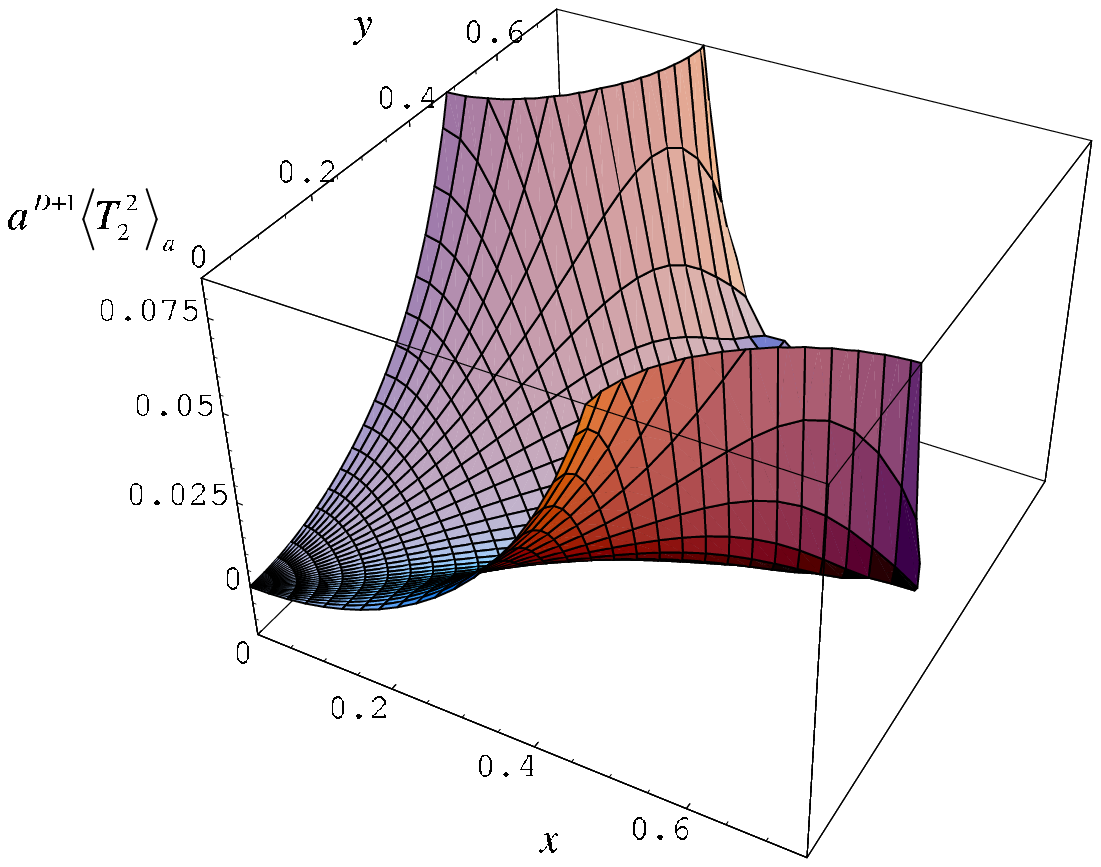, width=7cm, height=6cm}%
\end{tabular}%
\end{center}
\caption{The same as in figure \protect\ref{fig2} for the off-diagonal $%
{}^1_2$-component and the azimuthal stress.}
\label{fig3}
\end{figure}

For the angles $0<\phi <\phi _{0}$, by using the formula
\begin{equation}
\sum_{n=1}^{\infty }n^{D}e^{-\alpha n}\cos n\beta =\frac{(-1)^{D}}{2}\frac{%
d^{D}}{d\alpha ^{D}}\left( \frac{\sinh \alpha }{\cosh \alpha -\cos \beta }%
-1\right) ,  \label{formser1}
\end{equation}%
introducing a new integration variable $y=2\pi
(1-r/a)\sqrt{1+z^{2}}/\phi _{0}$, and expanding over $(1-r/a)$, one
finds that for $|\phi -\phi _{m}|\gg 1-r/a$ the leading contribution
of the term with $A_{i}^{(+)}(z)$ dominates and to the leading order
we find
\begin{equation}
\langle T_{i}^{i}\rangle _{a}\approx \frac{D(\xi -\xi _{c})\Gamma \left(
\frac{D+1}{2}\right) }{2^{D}\pi ^{(D+1)/2}(a-r)^{D+1}},\quad i=0,2,\ldots ,D.
\label{T00asra2}
\end{equation}%
This leading divergence coincides with the corresponding one for a
cylindrical surface of the radius $a$ (see, for instance, \cite{saharianI}).
For the other components to the leading order one has $\langle
T_{1}^{1}\rangle _{a}\sim \langle T_{2}^{1}\rangle _{a}\sim (a-r)^{-D}$. In
figures \ref{fig2} and \ref{fig3} we have plotted the dependences of the
boundary-induced vacuum expectation values for the components of the
energy-momentum tensor in the case of $D=3$ conformally coupled scalar field
as functions on coordinates $x=(r/a)\cos \phi $, $y=(r/a)\sin \phi $ for a
wedge with $\phi _{0}=\pi /2$. In particular, the corresponding
energy-density is negative everywhere and the vacuum forces determined by
the component $\langle T_{2}^{2}\rangle _{a}|_{\phi =\phi _{m}}$ are
attractive.

\section{Whightman function, the field square and the vacuum energy-momentum tensor
in the region II}

\label{sec:extregion}

Now we turn to the region outside the cylindrical shell (region II
in figure \ref{fig1}): $r>a$, $0\leq \phi \leq \phi_{0}$. The
corresponding eigenfunctions satisfying boundary conditions
(\ref{Dirbc}) are obtained from (\ref{eigfunccirc}) by the
replacement
\begin{equation}\label{replace}
    J_{qn}(\gamma r)\rightarrow  g_{qn}(\gamma r,\gamma a)\equiv
    J_{qn}(\gamma r)Y_{qn}(\gamma a)-
    J_{qn}(\gamma a)Y_{qn}(\gamma r),
\end{equation}
where $Y_{qn}(z)$ is the Neumann function. Now the spectrum for the
quantum number $\gamma $ is continuous. To determine the
corresponding normalization coefficient $\beta _{\alpha}$, we note
that as the normalization integral diverges in the limit $\gamma
=\gamma '$, the main contribution into the integral over radial
coordinate comes from the large values of $r$ when the Bessel
functions can be replaced by their asymptotics for large arguments.
The resulting integral is taken elementary and for the normalization
coefficient in the region $r>a$ one finds
\begin{equation}\label{norcoefext}
    \beta _{\alpha }^2=\frac{(2\pi )^{2-D}\gamma }{\phi _{0}\omega
    \left[ J_{qn}^{2}(\gamma a)+Y_{qn}^{2}(\gamma a)\right]}.
\end{equation}
Substituting the corresponding eigenfunctions into the mode sum
formula (\ref{vevWf}), the positive frequency Whightman function is
presented in the form
\begin{eqnarray}
\langle 0|\varphi (x)\varphi (x')|0\rangle &=& \frac{1}{\phi _{0}}
\int \frac{d^{N}{\mathbf{k}}}{(2\pi )^{N}}
e^{i{\mathbf{k}}({\mathbf{r}}_{\parallel }- {\mathbf{r}}'_{\parallel
})} \sum_{n=1}^{\infty }\sin (qn\phi )\sin (qn\phi ')
\nonumber \\
 && \times \int_{0}^{\infty }d\gamma \frac{\gamma g_{qn}(\gamma r,
 \gamma a)g_{qn}(\gamma r',\gamma a)}{J_{qn}
 ^2(\gamma a)+Y_{qn}^{2}(\gamma a)}\frac{\exp \left[i(t'-t)\sqrt{\gamma ^2
 +k^2}\right]}{\sqrt{\gamma ^2 +k^2}}. \label{Wfext0}
\end{eqnarray}
To find the part in the Wightman function induced by the presence of
the cylindrical shell we subtract from (\ref{Wfext0}) the
corresponding function for the wedge without a cylindrical shell,
given by (\ref{Wf00}). In order to evaluate the corresponding
difference we use the relation
\begin{equation}\label{relext}
\frac{g_{qn}(\gamma r,
 \gamma a)g_{qn}(\gamma r',\gamma a)}{J_{qn}
 ^2(\gamma a)+Y_{qn}^{2}(\gamma a)}-J_{qn}(\gamma r)J_{qn}(\gamma
 r')=-\frac{1}{2}\sum_{s=1}^{2}\frac{J_{qn}(\gamma a)}{H^{(s)}_{qn}(\gamma a)}
H^{(s)}_{qn}(\gamma r)H^{(s)}_{qn}(\gamma r'),
\end{equation}
where $H^{(s)}_{qn}(z)$, $s=1,2$ are the Hankel functions. This
allows to present the Wightman function in the form (\ref{Wf2}) with
the cylindrical shell induced part
\begin{eqnarray}
\langle \varphi (x)\varphi (x')\rangle _{a}&=& -\frac{1}{2\phi _{0}}
\int \frac{d^{N}{\mathbf{k}}}{(2\pi )^{N}}
e^{i{\mathbf{k}}({\mathbf{r}}_{\parallel }- {\mathbf{r}}'_{\parallel
})} \sum_{n=1}^{\infty }\sin (qn\phi )\sin (qn\phi ')
\nonumber \\
 && \times \sum_{s=1}^{2}\int_{0}^{\infty }d\gamma \gamma \frac{J_{qn}(\gamma a)}{H^{(s)}_{qn}(\gamma a)}
H^{(s)}_{qn}(\gamma r)H^{(s)}_{qn}(\gamma r')\frac{\exp
\left[i(t'-t)\sqrt{\gamma ^2
 +k^2}\right]}{\sqrt{\gamma ^2
 +k^2}}. \label{Wfext2}
\end{eqnarray}
On the complex plane $\gamma $ we can rotate the integration contour
on the right of this formula by the angle $\pi /2$ for $s=1$ and by
the angle $-\pi /2$ for $s=2$. The integrals over the segments
$(0,ik)$ and $(0,-ik)$ cancel out and after introducing the Bessel
modified functions we obtain
\begin{eqnarray}
\langle \varphi (x)\varphi (x^{\prime })\rangle _{a}
&=&-\frac{2}{\pi \phi
_{0}}\int \frac{d^{N}\mathbf{k}}{(2\pi )^{N}}e^{i\mathbf{k}(\mathbf{r}%
_{\parallel }-\mathbf{r}_{\parallel }^{\prime })}\int_{k}^{\infty }dz\frac{%
z\cosh \left[ (t-t^{\prime })\sqrt{z^{2}-k^{2}}\right]
}{\sqrt{z^{2}-k^{2}}}
\nonumber \\
&&\times \sum_{n=1}^{\infty }\sin (qn\phi )\sin (qn\phi ^{\prime
})K_{qn}(zr)K_{qn}(zr^{\prime })\frac{I_{qn}(za)}{K_{qn}(za)}.
\label{Wfa0ext}
\end{eqnarray}%
Taking the coincidence limit for the arguments, from this formula
we obtain the vacuum expectation value of the field square in the
region $r>a$:
\begin{equation}
\langle \varphi ^{2}\rangle _{a}=-\frac{2^{3-D}\pi
^{\frac{1-D}{2}}}{\Gamma \left( \frac{D-1}{2}\right) a^{D-1}\phi
_{0}}\sum_{n=1}^{\infty }\sin
^{2}(qn\phi )\int_{0}^{\infty }dz\,z^{D-2}\frac{I_{qn}(z)}{K_{qn}(z)}%
K_{qn}^{2}(zr/a).  \label{phi2a1ext}
\end{equation}%
Comparing with formulae (\ref{Wfa0}) and (\ref{phi2a1}), we see that
the parts induced by the cylindrical shell in the regions I and II
are obtained from each other by the interchange
$I_{qn}(z)\leftrightarrows K_{qn}(z)$.

As for the region I, vacuum expectation value (\ref{phi2a1ext})
diverges on the cylindrical surface. The leading term in the
corresponding asymptotic expansion near this surface is obtained
from that for the region I, formula (\ref{Phi2neara}), replacing
$(a-r)$ by $(r-a)$. For large distances from the cylindrical
surface, $r\gg a$, we introduce in (\ref{phi2a1ext}) a new
integration variable $y=zr/a$ and expand the integrand over $a/r$.
The main contribution comes from the $n=1$ term. By taking into
account the value for the standard integral involving the square of
the MacDonald function, to the leading order one finds
\begin{equation}\label{phi2larger}
\langle \varphi ^{2}\rangle _{a}\approx -\frac{\sin ^{2}(q\phi
)\Gamma \left( \frac{D-1}{2}+2q\right) \Gamma ^2\left(
\frac{D-1}{2}+q\right)}{\pi ^{\frac{D+1}{2}}a^{D-1}\Gamma \left(
D-1+2q\right) \Gamma ^2(q)} \left( \frac{a}{r}\right) ^{D-1+2q} .
\end{equation}

For the part in the vacuum energy-momentum tensor induced by the
cylindrical surface in the region $r>a$, from (\ref{vevEMTWf}),
(\ref{Wfa0ext}), (\ref{phi2a1ext}) one has the following formulae
\begin{eqnarray}
\langle T_{i}^{i}\rangle _{a} &=&\frac{(4\pi
)^{-\frac{D-1}{2}}}{\Gamma \left( \frac{D-1}{2}\right) a^{D+1}\phi
_{0}}\sum_{n=1}^{\infty
}\int_{0}^{\infty }dzz^{D}\frac{I_{qn}(z)}{K_{qn}(z)}  \nonumber \\
&&\times \left\{
a_{i,qn}^{(+)}[K_{qn}(y)]-a_{i,qn}^{(-)}[K_{qn}(y)]\cos (2qn\phi
)\right\}
,\quad y=\frac{zr}{a},  \label{Tiiaext} \\
\langle T_{2}^{1}\rangle _{a} &=&\frac{2(4\pi
)^{-\frac{D-1}{2}}}{\Gamma \left( \frac{D-1}{2}\right) a^{D}\phi
_{0}}\sum_{n=1}^{\infty }qn\sin
(2qn\phi )\int_{0}^{\infty }dz\,z^{D-1}\frac{I_{qn}(z)}{K_{qn}(z)}  \nonumber \\
&&\times K_{qn}(y)\left[ \frac{2\xi }{y}K_{qn}(y)+(1-4\xi
)K_{qn}^{\prime }(y)\right] ,  \label{Tiia21ext}
\end{eqnarray}%
with the functions $a_{i,qn}^{(\pm )}[g(y)]$ defined by
(\ref{ajpm})-(\ref{ajpm2}). By the way similar to that used above
for the vacuum expectation value of the field square, it can be seen
that for large distances from the cylindrical surface, $r\gg a$, the
main contribution comes from the term with $n=1$ and the components
of the induced energy-momentum tensor behave as $\langle
T_{i}^{i}\rangle _{a} \sim (a/r)^{D+1+2q}$, $\langle
T_{2}^{1}\rangle _{a}\sim (a/r)^{D+2q}$. Near the cylindrical
surface the leading terms of the asymptotic expansions for the
components of the energy-momentum tensor in the exterior region are
obtained from the corresponding formulae for the interior region,
formulae (\ref{T00asra1}), (\ref{T00asra2}), replacing $(a-r)$ by
$(r-a)$. As for the interior region, the vacuum forces acting on the
wedge sides due to the presence of the cylindrical shell are
attractive and the corresponding energy density is negative for both
minimally and conformally coupled scalars.

\section{Limiting case}

\label{sec:limit}

\begin{figure}[tbph]
\begin{center}
\epsfig{figure=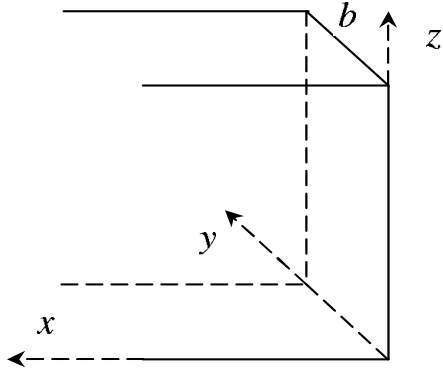, width=6cm, height=5cm}
\end{center}
\caption{Geometry of two parallel plates with the interplate distance $b$
perpendicularly intersected by the plate at $x=0$.}
\label{fig4}
\end{figure}

In this section we consider a limiting case of previous results,
obtained in the limit $\phi _{0}\rightarrow 0$, $r,a\rightarrow
\infty $, assuming that $a-r$ and $a\phi _{0}\equiv b$ are fixed.
This limit corresponds to the geometry of two parallel plates
separated by a distance $b$, perpendicularly intersected by the
third plate (see figure \ref{fig4}). The vacuum expectation values
of the energy-momentum tensor for this geometry of boundaries are
investigated in \cite{Acto96} for special cases of minimally and
conformally coupled massless scalar fields. Formulae presented below
in this section generalize this results for the case of an arbitrary
coupling and give an alternative representation of the vacuum
energy-momentum tensor. We introduce rectangular coordinates
$(x^{\prime 1},x^{\prime 2},\ldots ,x^{\prime D})=(x,y,z_{1},\ldots
,z_{N})$ with the relations $x=a-r$, $y=r\phi $ in the limit under
consideration. Below the components of the tensors in these
coordinates we will denote by primes. In this limit, from the
quantities corresponding to the wedge without a cylindrical surface
we obtain the vacuum densities in the region between two parallel
planes. These quantities are well-investigated in literature (see
\cite{Rome02} for the case of general mixed boundary conditions) and
below we will consider the additional part induced by the presence
of the intersecting plate at $x=0$. The corresponding vacuum
expectation values are obtained from the expectation values $\langle
\cdots \rangle _{a}$ investigated in previous section. For this we
note that in the limit under consideration one has $q=\pi /\phi
_{0}\rightarrow \infty $, and the order of the Bessel modified
functions in formulae (\ref{phi2a1}), (\ref{Tiia}), (\ref{Tiia21})
tends to infinity. Introducing a new integration variable
$z\rightarrow qnz$, we can replace these functions by their uniform
asymptotic expansions for large values of the order. The vacuum
expectation value of the field square is presented in
the form%
\begin{equation}
\langle 0|\varphi ^{2}|0\rangle =\langle \varphi ^{2}\rangle ^{(0)}+\langle
\varphi ^{2}\rangle ^{(1)},  \label{phi2lim}
\end{equation}%
where $\langle \varphi ^{2}\rangle ^{(0)}$ is the vacuum expectation value
for two parallel planes, and
\begin{equation}
\langle \varphi ^{2}\rangle ^{(1)}=-\frac{x^{1-\frac{D}{2}}}{2^{D-2}\pi b^{%
\frac{D}{2}}}\sum_{n=1}^{\infty }n^{\frac{D}{2}-1}\sin ^{2}(nv)K_{D/2-1}(nu)
\label{phi2lim1}
\end{equation}%
is induced by the presence of the plate at $x=0$. In (\ref{phi2lim1}) we use
the notations%
\begin{equation}
u=2\pi x/b,\quad v=\pi y/b.  \label{unvn}
\end{equation}
Formula (\ref{phi2lim1}) can also be directly obtained by using the
mode sum formula (\ref{vevWf}) with the eigenfunctions for the
boundary geometry under consideration:
\begin{equation}\label{phialflim}
    \varphi _{\alpha }=\frac{2\sin (k_1 x)}{\sqrt{(2\pi )^{D-1}b \omega
    }}\sin (\pi n y/b) \exp(i{\mathbf{kr}}_{\parallel }-i\omega t),
\end{equation}
where $0<k_1<\infty $, $n=1,2,\ldots $, and $\omega
=\sqrt{k^2+k_1^2+(\pi n/b)^2}$.

The representation similar to (\ref{phi2lim}) takes place for the components
of the energy-momentum tensor:%
\begin{equation}
\langle 0|T_{k}^{\prime i}|0\rangle =\langle T_{k}^{\prime i}\rangle
^{(0)}+\langle T_{k}^{\prime i}\rangle ^{(1)},  \label{Tiklim}
\end{equation}%
where the expectation values $\langle T_{k}^{\prime i}\rangle ^{(1)}$ are
induced by the intersecting plate at $x=0$. They are related to the
quantities investigated in previous section by formulae%
\begin{equation}
\langle T_{i}^{\prime i}\rangle ^{(1)}=\lim \,\langle T_{i}^{i}\rangle
_{a},\quad \langle T_{2}^{\prime 1}\rangle ^{(1)}=-\lim \frac{1}{a}\langle
T_{2}^{1}\rangle _{a},  \label{limtrans}
\end{equation}%
with $\lim $ corresponding to the limit $a\rightarrow \infty $, $\phi
_{0}\rightarrow 0$ for fixed $a-r$ and $a\phi _{0}$. Using the formulae for $%
\langle T_{i}^{k}\rangle _{a}$ for the induced quantities one finds (no
summation over $i$, $i=0,3,\ldots ,D$)%
\begin{eqnarray}
\langle T_{i}^{\prime i}\rangle ^{(1)} &=&\frac{\pi x^{1-\frac{D}{2}}}{%
2^{D-2}b^{\frac{D}{2}+2}}\sum_{n=1}^{\infty }n^{\frac{D}{2}+1}\left\{ \frac{%
(4\xi -1)(D-1)}{nu}\sin ^{2}(nv)K_{D/2}(nu)\right.   \nonumber \\
&&+\left. \left[ 2\xi -\frac{1}{2}+\frac{\sin ^{2}(nv)}{D-1}\right]
K_{D/2-1}(nu)\right\} ,  \label{Tiilim} \\
\langle T_{1}^{\prime 1}\rangle ^{(1)} &=&\frac{\pi (4\xi -1)x^{1-\frac{D}{2}%
}}{2^{D-1}b^{\frac{D}{2}+2}}\sum_{n=1}^{\infty }n^{\frac{D}{2}+1}\cos
(2nv)K_{D/2-1}(nu),  \label{T11lim} \\
\langle T_{2}^{\prime 1}\rangle ^{(1)} &=&\frac{\pi (4\xi -1)x^{1-\frac{D}{2}%
}}{2^{D-1}b^{\frac{D}{2}+2}}\sum_{n=1}^{\infty }n^{\frac{D}{2}+1}\sin
(2nv)K_{D/2}(nu),  \label{T21lim} \\
\langle T_{2}^{\prime 2}\rangle ^{(1)} &=&\frac{\pi x^{1-\frac{D}{2}}}{%
2^{D-2}b^{\frac{D}{2}+2}}\sum_{n=1}^{\infty }n^{\frac{D}{2}+1}\left\{ \frac{%
(4\xi -1)(D-1)}{nu}\sin ^{2}(nv)K_{D/2}(nu)\right.   \nonumber \\
&&+\left. \left[ \frac{1}{2}+(4\xi -1)\sin ^{2}(nv)\right]
K_{D/2-1}(nu)\right\} ,  \label{T22lim}
\end{eqnarray}%
with the notations from (\ref{unvn}). Due to the presence of the
MacDonald function in this formulae, the series are exponentially
convergent. Note that in the alternative formulae for the
energy-momentum tensor for minimally and conformally coupled scalar
fields, given in \cite{Acto96}, the convergence of the series is
power-law.

For $x>0$ quantities (\ref{Tiilim})-(\ref{T22lim}) are finite
on the plates $y=0,b$:%
\begin{equation}
\langle T_{i}^{\prime i}\rangle ^{(1)}|_{y=0,b}=\frac{\pi x^{1-\frac{D}{2}%
}A_{i}}{2^{D-1}b^{\frac{D}{2}+2}}\sum_{n=1}^{\infty }n^{\frac{D}{2}%
+1}K_{D/2-1}(nu),  \label{Tiilimonplate}
\end{equation}%
and $\langle T_{2}^{\prime 1}\rangle ^{(1)}|_{y=0,b}=0$. In particular,
additional vacuum forces acting on the plates $y=0,b$ due to the presence of
the plate at $x=0$ are determined by the component $\langle T_{2}^{\prime
2}\rangle ^{(1)}$ and are attractive. For large distances from the plate $x=0
$, by using the asymptotic formula for the modified Bessel function, we can
see that the main contribution comes from $n=1$ term and to the leading
order we find%
\begin{equation}
\langle T_{i}^{\prime k}\rangle ^{(1)}\sim u^{\frac{1-D}{2}}e^{-u},\quad
u\gg 1,  \label{Tiklargeu}
\end{equation}%
with $u$ defined by relation (\ref{unvn}). In this limit the vacuum
expectation values induced by the plate $x=0$ are exponentially suppressed.

\section{Conclusion}

\label{sec:Conc}

We have investigated the Wightman function, vacuum expectation
values of the field square and the energy-momentum tensor for a
massless scalar field with a general curvature coupling parameter
inside a wedge with a coaxial cylindrical boundary. We have
assumed Dirichlet boundary conditions on the bounding surfaces.
The generalization of the corresponding results for other boundary
conditions is straightforward. The application of the generalized
Abel-Plana summation formula for the series over the zeros of the
Bessel function allowed to extract from the expectation values the
parts due to the wedge without a cylindrical boundary and to
present the additional parts induced by this boundary in terms of
exponentially convergent integrals. The vacuum densities for the
geometry of a wedge without a cylindrical boundary are considered
in section \ref{sec:Wedgecylabs}. We have derived formulae for the
renormalized vacuum expectation values of the field square and the
energy-momentum tensor, formulae (\ref{phi2w}), (\ref{Tiiw}), (\ref{Tiiw21}%
). These formulae can be further simplified for the important special case $%
D=3$ [formulae (\ref{phi2D3}), (\ref{TikD3})-(\ref{TikD322})]. For a
conformally coupled scalar the energy-momentum tensor is diagonal and does
not depend on the angular variable $\phi $. The corresponding vacuum forces
acting on the wedge sides are attractive for $\phi _{0}<\pi $ and are
repulsive for $\phi _{0}>\pi $. The previous investigations of the problem
are restricted to this case.

In section \ref{sec:Wedgecyl} we have investigated the additional
expectation values for the field square and the energy-momentum
tensor induced by the presence of the cylindrical surface. The
field square is given by formula (\ref{phi2a1}) and vanishes on
the wedge sides $\phi =\phi _{m}$ for all points away from the
cylindrical surface. The energy-momentum tensor induced by the
cylindrical surface is
non-diagonal and the corresponding components are determined by formula (\ref%
{Tiia}), (\ref{Tiia21}). The off-diagonal component vanishes on the wedge
sides. The additional vacuum forces acting on the wedge sides due to the
presence of the cylindrical surface are determined by the $_{2}^{2}$%
-component of the corresponding stress and are attractive for all values $%
\phi _{0}$. On the wedge sides the corresponding vacuum stresses in
the directions parallel to the wedge sides are isotropic and the
energy density is negative for both minimally and conformally
coupled scalars. The vacuum expectation values diverge on the
cylindrical surface. For the points with $|\phi -\phi _{m}|\gg
1-r/a$ the leading divergences are the same as those for a
cylindrical surface without the wedge. As an illustration of the
general results, we have plotted in figures \ref{fig2} and
\ref{fig3} the components of the vacuum energy-momentum tensor
induced by the cylindrical surface in the case of a conformally
coupled $D=3$ scalar field for a wedge with the opening angle $\phi
_{0}=\pi /2$. The Wightman function, vacuum expectation values of
the field square and the energy-momentum tensor in the region
outside the cylindrical shell are investigated in section
\ref{sec:extregion}. The formulae for these quantities differ from
the corresponding formulae for the interior region by the
interchange $I_{qn}(z)\leftrightarrows K_{qn}(z) $. For large
distances from the cylindrical surface, $r\gg a$, the vacuum
expectation values behave as $(a/r)^{D-1+2q}$ for the field square
and as $(a/r)^{D+1+2q}$ for the diagonal components of the
energy-momentum tensor. As in the case of interior region, the
vacuum forces acting on the wedge sides due to the presence of the
cylindrical shell are attractive. In section \ref{sec:limit} we have
considered a limiting case $\phi _{0}\rightarrow 0$, $r,a\rightarrow
\infty $, assuming that $a-r$ and $a\phi _{0}$ are fixed. This limit
corresponds to the geometry of two parallel plates perpendicularly
intersected by the third plate (see figure \ref{fig4}) and has been
investigated previously in \cite{Acto96} for special cases of
minimally and conformally coupled scalar fields. For general values
of the curvature coupling parameter the additional vacuum
expectation values induced by the intersecting plate are given by
formulae (\ref{phi2lim1}), (\ref{Tiilim})-(\ref{T22lim}). In
particular, the corresponding energy-momentum tensor is diagonal on
the parallel plates with the components (\ref{Tiilimonplate}) and
the additional vacuum forces due to the presence of the intersecting
plate are attractive. Vacuum densities induced by this plate are
exponentially suppressed for large distances.

The generalization of the results obtained here for the Neumann,
or more general Robin boundary conditions is straightforward. For
instance, for the Neumann case in the expressions
(\ref{eigfunccirc}) of the eigenfunctions the function $\cos
(qn\phi )$ stands instead of $\sin (qn\phi )$ and the quantum
number $n$ takes the values $0,1,2,\ldots $. In the case with a
cylindrical boundary now the eigenvalues for $\gamma a$ are zeros
for the derivative of the Bessel function. The formula to sum the
series over these zeros can be
taken from \cite{Saha87}.  Now, in formulae (\ref{phi2a1}), (\ref{Tiia}), (%
\ref{Tiia21}) for the vacuum expectation values, instead of the ratio $%
K_{qn}(z)/I_{qn}(z)$ the ratio of derivatives, $K_{qn}^{\prime
}(z)/I_{qn}^{\prime }(z)$, will stand. Note that in this paper we have
considered boundary induced vacuum densities which are finite away from the
boundaries. As it has been mentioned in Ref. \cite{Grah03}, the same results
will be obtained in the model where instead of externally imposed boundary
condition the fluctuating field is coupled to a smooth background potential
that implements the boundary condition in a certain limit \cite{Grah02}.

\section*{Acknowledgments}

The work of AAS was supported by ANSEF Grant No. 05-PS-hepth-89-70
and in part by the Armenian Ministry of Education and Science
Grant No. 0124.

\end{document}